%% file: main.tex
\documentclass[11pt,letterpaper]{article}

\usepackage[T1]{fontenc}
\usepackage{lmodern}

\usepackage[utf8]{inputenc} % allow utf-8 input
\usepackage[T1]{fontenc}    % use 8-bit T1 fonts
\usepackage[hidelinks]{hyperref}       % hyperlinks
\usepackage{url}            % simple URL typesetting
\usepackage{booktabs}       % professional-quality tables
\usepackage{amsfonts}       % blackboard math symbols
\usepackage{nicefrac}       % compact symbols for 1/2, etc.
\usepackage{microtype}      % microtypography

\usepackage{notoccite}

\usepackage{amsmath} % assumes amsmath package installed
\usepackage{amssymb}  % assumes amsmath package installed
\usepackage{color,xspace}
\usepackage[hidelinks]{hyperref}
\usepackage{graphicx}
\usepackage{cite}
\usepackage{subcaption}
\usepackage{booktabs}

\usepackage[margin=1.0in]{geometry}

\allowdisplaybreaks

\newcommand{\DDelta}{\mathbf{\Delta}}
\newcommand{\xx}{\mathbf{x}}
\newcommand{\uu}{\mathbf{u}}
\newcommand{\ww}{\mathbf{w}}
\newcommand{\bfeta}{\boldsymbol{\eta}}
\newcommand{\dd}{\mathbf{d}}
\newcommand{\sA}{\mathbf{A}}
\newcommand{\sB}{\mathbf{B}}

\newcommand{\Phix}{\mathbf{\Phi}_x}
\newcommand{\Phiu}{\mathbf{\Phi}_u}
\newcommand{\KK}{\mathbf{K}}

\newcommand{\tildePhix}{\widetilde{\mathbf{\Phi}}_x}
\newcommand{\tildePhiu}{\widetilde{\mathbf{\Phi}}_u}
\newcommand{\tildeww}{\widetilde{\mathbf{w}}}

\newcommand{\conv}{\textrm{Co}}
\newcommand{\bfSigma}{\mathbf{\Sigma}}
\newcommand{\facet}{\textrm{facet}}

\newcommand{\diag}{\textrm{diag}}
\newcommand{\blkdiag}{\textrm{blkdiag}}

\newcommand{\vertex}{\textrm{Vert}}

\newcommand{\bd}{\mathbf{d}}
\newcommand{\ddelta}{\boldsymbol{\delta}}
\newcommand{\hh}{\mathbf{h}}
\newcommand{\tildexx}{\widetilde{\mathbf{x}}}
\newcommand{\CC}{\mathbf{C}}
\newcommand{\vv}{\mathbf{v}}
\allowdisplaybreaks

\input defs.tex

\title{{\LARGE \bf Robust Model Predictive Control of Time-Delay Systems through System Level Synthesis}}

\author{Shaoru Chen, Ning-Yuan Li, Victor M. Preciado, Nikolai Matni 
\thanks{Shaoru Chen, Ning-Yuan Li, Victor M. Preciado, Nikolai Matni are with the Department of Electrical and Systems Engineering, University of Pennsylvania, Philadelphia, PA, 19104, USA (e-mail: \{srchen, ny0221, preciado, nmatni\}@seas.upenn.edu). Nikolai Matni is funded by NSF awards CPS-2038873, CAREER award ECCS-2045834, and a Google Research Scholar award. }
}
 
\date{}

\begin{document}
\pagestyle{plain}
%\pagenumbering{arabic}
\maketitle

\begin{abstract}
\input{abstract}
\end{abstract}

\section{Introduction}
\input{introduction}

\section{Problem Formulation}
\input{formulation}

\section{SLS-based Robust MPC: Non-Delay Case}
\input{SLSMPC}

\section{Effects of Time Delay}
\input{delaySLSMPC}

\section{SLS-based Time-delay Robust MPC}
\input{SLSMethod}

\section{Simulation}
\input{new_simulation}

\section{Conclusion}
\input{conclusion}

\appendix

\input{new_appendix}

\bibliographystyle{ieeetr}
\bibliography{Refs}

% 1. need to mention closed-loop feasibility and stability 
% 2. 

\end{document}

%% file: defs.tex
\newtheorem{theorem}{Theorem}

\newtheorem{remark}{Remark}
\newtheorem{example}{Example}

\newtheorem{problem}{Problem}

%% file: abstract.tex
% abstract
We present a robust model predictive control method (MPC) for discrete-time linear time-delayed systems with state and control input constraints. The system is subject to both polytopic model uncertainty and additive disturbances. In the proposed method, a time-varying feedback control policy is optimized such that the robust satisfaction of all constraints for the closed-loop system is guaranteed. By encoding the effects of the delayed states and inputs into the feedback policy, we solve the robust optimal control problem in MPC using System Level Synthesis which results in a convex quadratic program that jointly conducts uncertainty over-approximation and robust controller synthesis. Notably, the number of variables in the quadratic program is independent of the delay horizon. The effectiveness and scalability of our proposed method are demonstrated numerically. 

%% file: introduction.tex
% introduction
Time-delay systems appear in many applications such as chemical process control, communication networks, and aircraft control since they are suitable to model the non-instantaneous behavior of physical processes and capture the time needed to transport information. However, controlling time-delay systems is challenging since time delay can seriously degrade the performance and induce instability of the closed-loop system. The control task becomes even more challenging when state and input constraints are considered, the system dynamics are uncertain, and process/measurement noise needs to be taken into account in many real-world applications. 

Robust model predictive control (MPC) is promising to address the above issues with closed-loop safety (i.e., all state and control input constraints are satisfied) and stability guarantees. In robust MPC, at each time instant, a finite-horizon robust optimal control problem (OCP) is solved to synthesize a robust control input. There is a rich body of work in robust MPC for uncertain linear systems without time delay where different controller parameterization and robust OCP formulations are proposed, such as linear matrix inequalities (LMI)~\cite{kothare1996robust}, tubes~\cite{langson2004robust}, state/disturbance feedback controllers~\cite{goulart2006optimization}, and System Level Synthesis (SLS)~\cite{sieber2021system, chen2022robust}. These methods have distinctive complexity-conservatism trade-offs and robust MPC remains an active area of research.

Although we can transform a discrete-time delayed system into a non-delay one through state and input augmentation~\cite[Chapter 6]{fridman2014introduction}, this approach can easily lead to a high-dimensional system which is challenging to handle for any robust MPC method. Due to the difficulty of handling time delays, available robust MPC methods for discrete-time delayed systems are as diverse as in the non-delay case. Indeed, they rely mostly on LMI formulations~\cite{shi2009delay, bououden2016constrained, ding2007constrained} to synthesize a locally stabilizing linear time-invariant state feedback controller. This approach can be conservative since it only searches over ellipsoidal robust invariant sets. With a fixed locally stabilizing controller, the authors in~\cite{laraba2017linear, olaru2008predictive} apply an iterative algorithm to find the polytopic maximal robust invariant set which can be used as the terminal set in MPC to guarantee closed-loop stability. However, in these approaches, the robust OCP is solved conservatively over open-loop control inputs rather than feedback policies. 

\textbf{Contributions}: In this paper, we propose a novel robust state-feedback controller design for robust MPC of time-delay systems subject to polytopic model uncertainty and additive disturbances. Our method is built on robust SLS MPC~\cite{chen2022robust} for non-delay systems which over-approximates the perturbations to the nominal dynamics by a surrogate additive disturbance, but we extend~\cite{chen2022robust} to time-delay systems in a non-trivial way. Due to the delay, the surrogate disturbance over-approximation in SLS MPC can be overly conservative since the state and input delays incur an off-set to the system perturbations. To address this issue, in this work we design a time-varying feedback controller that integrates the time-delay effects to solve the robust OCP. Furthermore, the proposed method solves a convex quadratic program (QP) for controller synthesis. Since we do not apply any state or input augmentation, the number of optimization variables in the QP is independent of the delay horizon, making our method scalable to systems with large delay horizons. The effectiveness and scalability of our method is demonstrated numerically.

%This paper is structured as follows. Robust MPC of uncertain time-delay systems is formulated in Section~\ref{sec:formulation}. We present SLS-based robust MPC for non-delay systems in Section~\ref{sec:SLS_MPC}. We analyze the challenges brought by the time delay and present our controller design in Section~\ref{sec:delay_effects}. In Section~\ref{sec:time_delay_MPC} we formulate a tractable robust OCP using SLS. Section~\ref{sec:simulation} shows an illustrative example and Section~\ref{sec:conclusion} concludes the paper.

\textbf{Notation }  We represent a linear, causal operator $\mathbf{R}$ defined over a horizon of $T$ by the block-lower-triangular matrix
\begin{equation} \label{eq:BLT}
\mathbf{R} = \begin{bmatrix}
R^{0,0} & \ & \ & \ \\
R^{1,1} & R^{1,0} & \ & \ \\
\vdots & \ddots & \ddots & \ \\
R^{T,T} & \cdots &R^{T,1} & R^{T,0}
\end{bmatrix}
\end{equation} 
where $R^{i,j} \in \mathbb{R}^{p \times q}$ is a matrix of compatible dimension. The set of such matrices is denoted by $\mathcal{L}_{TV}^{T, p \times q}$ and we will drop the superscript $T, p \times q$ when it is clear from the context. We refer to a block matrix in a block-lower-triangular matrix $\mathbf{R}$ using its superscripts shown in~\eqref{eq:BLT}. Let $\mathbf{R}(i,:)$ denote the $i$-th block row of $\mathbf{R}$, and $\mathbf{R}(:,j)$ denote the $j$-th block column of $\mathbf{R}$, both indexing from $0$. For a vector $d \in \mathbb{R}^n$, $S = \diag(d)$ denotes a $n \times n$ dimensional diagonal matrix with $d$ being the diagonal elements. The notation $x_{0:T}$ is shorthand for the set $\{x_0, \cdots, x_T\}$. 

%% file: formulation.tex
% formulation
\label{sec:formulation}
Consider the following discrete-time linear system with time delay:
\begin{equation}\label{eq:delay_system}
x(k+1) = \sum_{i=0}^{n_a} A_i x(k-i) + \sum_{j = 0}^{n_b} B_j u(k-j) + w(k)
\end{equation}
where $x(k) \in \mathbb{R}^{n_x}$ is the state state, $u(k) \in \mathbb{R}^{n_u}$ is the control input, $w(k)\in\mathbb{R}^{n_x}$ is the additive disturbance at time $k$, and $n_a, n_b \geq 0$ denote the horizon of delay in states and control inputs, respectively. The dynamics $(A_{0:n_a}, B_{0:n_b})$ of the time-delay system is uncertain and is represented by 
\begin{equation} \label{eq:delayed_matrices}
A_i = \hat{A}_i + \Delta_{A,i}, 0 \leq i \leq n_a, B_j = \hat{B}_j + \Delta_{B, j}, 0 \leq j \leq n_b,
\end{equation}
where $\hat{A}_i, \hat{B}_j$ denote the nominal dynamics and the model uncertainty $(\Delta_{A,0:n_a}, \Delta_{B,0:n_b})$ belongs to a polytopic set $\mathcal{P}$:
\begin{equation}\label{eq:poly_uncertainty}
\begin{aligned}
(\Delta_{A,0:n_a}, \Delta_{B,0:n_b}) \in \mathcal{P} := \conv\{ (\Delta_{A,0:n_a}^1, \Delta_{B,0:n_b}^1), \cdots (\Delta_{A,0:n_a}^M, \Delta_{B,0:n_b}^M) \}
\end{aligned}
\end{equation}
where $\conv$ denotes the convex hull of $M$ vertices $(\Delta_{A,0:n_a}^\ell, \Delta_{B,0:n_b}^\ell)$. The disturbance $w(k)$ is assumed to be norm-bounded, i.e., $w(k) \in \mathcal{W} := \{w \in \mathbb{R}^{n_x} \mid \lVert w \rVert_\infty \leq \sigma_w\}$. The initial conditions $x(0), \cdots, x(-n_a)$ and $u(-1), \cdots, u(-n_b)$ of system~\eqref{eq:delay_system} are given. We allow the model uncertainty parameters $(\Delta_{A,0:n_a}, \Delta_{B,0:n_b})$ to be time-varying as long as they satisfy~\eqref{eq:poly_uncertainty}.

Our goal is to design a robust MPC controller for the time-delay system~\eqref{eq:delay_system} such that the state and control input constraints
\begin{equation}
	\begin{aligned}
	&\mathcal{X} = \{x \in \mathbb{R}^{n_x} \mid F_\mathcal{X} x \leq b_\mathcal{X}\},\ \mathcal{U} = \{u \in \mathbb{R}^{n_u} \mid F_\mathcal{U} x \leq b_\mathcal{U}\},
	\end{aligned}
\end{equation}
are satisfied robustly for the closed-loop system. The focus of this paper is on efficiently solving the robust OCP in each iteration of MPC which is formally stated as follows.
\begin{problem}
Solve the following finite time constrained robust OCP with horizon $T$:
\begin{equation} \label{eq:robustOCP}
	\begin{aligned}
	\underset{\pi}{\textrm{minimize}}  &\quad  J_T(x_{0:T}, u_{0:T-1}) \\
	\textrm{subject to} & \quad x_{t+1} = \sum_{i=0}^{n_a} (\hat{A}_i + \Delta_{A,i}) x_{t-i} + \sum_{j = 0}^{n_b} (\hat{B}_j + \Delta_{B, j}) u_{t-j} + w_t, \\
	& \quad u_{t} = \pi_t(x_{-n_a:t}, u_{-n_b:t-1}), \\
	& \quad x_t \in \mathcal{X}, u_t \in \mathcal{U}, x_T \in \mathcal{X}_T,\\
	& \quad \forall (\Delta_{A,0:n_a}, \Delta_{B,0:n_b}) \in \mathcal{P}, \forall w_t \in \mathcal{W}, \\
	& \quad t = 0, \cdots, T-1, \\
	& \quad x_{-n_a:0}, u_{-n_b:-1} \textrm{ are known.}
	\end{aligned}
\end{equation}
where the search is over robust causal feedback policies $\pi = \pi_{0:t}$ with known delayed states and control inputs $x_{-n_a:0}, u_{-n_b:-1}$. The cost function $J_T(x_{0:T}, u_{0:T-1})$ is application-specific but is assumed convex in its arguments. The terminal set $\mathcal{X}_T$ is a polytope given by 
\begin{equation} \label{eq:terminal_set}
\mathcal{X}_T = \{x \in \mathbb{R}^{n_x} \mid F_{\mathcal{X}_T} x \leq b_{\mathcal{X}_T}\}.
\end{equation}
We assume that $\mathcal{X}, \mathcal{U}, \mathcal{X}_T$ are compact and contain the origin in the interior~\footnote{Our proposed method allows imposing different polyhedral constraints on $x_t$ at different time instants. For example, the terminal constraint $\mathcal{X}_T$ can be imposed on $x_{T-n_a:T}$ rather than on $x_T$ only. }. 
\end{problem}

At time $k$, robust MPC solves problem~\eqref{eq:robustOCP} with $x_{-i} = x(k-i)$ and $u_{-j} = u(k-j)$ for $0 \leq i \leq n_a$, $1 \leq j \leq n_b$. The terminal constraint~\eqref{eq:terminal_set} is often used to guarantee closed-loop stability of MPC and can be chosen as a robust forward invariant set for a locally stabilizing controller~\cite{laraba2017linear}. For the robust OCP~\eqref{eq:robustOCP}, it is required to choose the horizon $T$ greater than $n_a$ and $n_b$ in order to fully evaluate the effects of the predicted control inputs. 

%For computational consideration, we choose the cost function $J_T$ in~\eqref{eq:robustOCP} as the nominal cost function \shaoru{make weight matrices time varying}:
%\begin{equation}
%\begin{aligned}
%J_T &= \sum_{t = 0}^{T - 1} (\hat{x}_t^\top Q \hat{x}_t + u_t^\top R u_t ) + \hat{x}_T^\top Q_T \hat{x}_T \\ 
%\textrm{s.t. } & \hat{x}_{t+1} = \sum_{i=0}^{n_a} \hat{A}_i \hat{x}_{t-i} + \sum_{j=0}^{n_b} \hat{B}_j u_{t-j} + w_t, \\
%& u_t = \pi_t(\hat{x}_{-n_a:t}, u_{-n_b:t-1}), t = 0, \cdots, T-1,
%\end{aligned}
%\end{equation}
%where the initial conditions $x_{-n_a:0}, u_{-n_b:-1}$ are the same as in~\eqref{eq:robustOCP}. 

%Similar to robust MPC of non-delay uncertain systems, to make problem~\eqref{eq:robustOCP} tractable we need to apply a finite dimensional parameterization of the feedback policy $\pi$. Exactly optimizing over $\pi$ is often intractable and an efficient algorithm to find a feasible and sub-optimal solution to the robust OCP~\eqref{eq:robustOCP} is needed. 

For the time-delay system~\eqref{eq:delay_system}, a central problem is how to handle the effects of delay when solving the robust OCP~\eqref{eq:robustOCP} with a finite-dimensional, parameterized feedback policy $\pi$. In this work, we design $\pi$ as a time-varying feedback policy which operates on both the states $x_{0:T}$ and the transformed delayed states $x_{-n_a:-1}$ and inputs $u_{-n_b:-1}$ whose values are known to us. This allows us to over-approximate the effects of uncertainty in system~\eqref{eq:delay_system} using SLS with minimal conservatism while maintaining the convexity of our proposed robust OCP. We provide background on SLS-based robust MPC developed for non-delay systems in Section~\ref{sec:SLS_MPC} and present our method for the time-delay system~\eqref{eq:delay_system} in Section~\ref{sec:delay_effects} and~\ref{sec:time_delay_MPC}.

%% file: SLSMPC.tex
% robust SLS MPC
\label{sec:SLS_MPC}
Before we approach the time-delay system, in this section, we introduce the methodology of robust SLS MPC~\cite{chen2022robust} for non-delay systems. The main idea of robust SLS MPC is to over-approximate the effects of model uncertainty and additive disturbances in the system dynamics by a surrogate filtered disturbance. Then, by SLS we can jointly search over robust linear state feedback controllers and uncertainty over-approximation parameters in the space of closed-loop system responses by a convex program. 

\subsection{Finite-horizon System Level Synthesis}
Consider the following linear time-varying (LTV) system 
\begin{equation}~\label{eq:LTV_system}
x_{t+1} = \hat{A}_t x_t + \hat{B}_t u_t + \eta_t, \ t \geq 0,
\end{equation}
where $\eta_t \in \mathbb{R}^{n_x}$ is the perturbation to the nominal dynamics. With a slight abuse of notation, in this subsection the matrices $\hat{A}_t, \hat{B}_t$ denote the nominal dynamics of~\eqref{eq:LTV_system} rather than the delayed dynamics matrices in~\eqref{eq:delayed_matrices}.

To describe the behavior of the LTV system~\eqref{eq:LTV_system} 
over a finite horizon $T$, we first introduce the following compact notation
\begin{equation} \label{eq:x_u}
\begin{aligned}
&\xx = [x_0^\top \ \cdots \ x_T^\top]^\top, \quad \uu = [u_0^\top \ \cdots \ u_T^\top]^\top, \quad \bfeta = [x_0^\top \ \eta_0^\top \ \cdots \ \eta_{T-1}^\top]^\top,
\end{aligned}
\end{equation}
where $\xx, \uu, \bfeta$ stack the relevant states, control inputs, and perturbations over horizon $T$ and can be interpreted as finite horizon signals. Note that the first component of $\bfeta$ is set as the initial state $x_0$. Then, the system dynamics~\eqref{eq:LTV_system} over horizon $T$ can be written as
\begin{equation} \label{eq:open_dyn}
\xx = Z \hat{\sA} \xx + Z \hat{\sB} \uu + \bfeta
\end{equation} 
where $Z \in \mathcal{L}_{TV}^{T, n_x\times n_x}$ is a block down-shifting operator with identity matrices filling the first block sub-diagonal and zeros everywhere else, and 
\begin{equation}
\begin{aligned}
&\mathbf{\hat{A}} = \blkdiag(\hat{A}_0, \cdots, \hat{A}_{T-1}, 0), \quad \mathbf{\hat{B}} = \blkdiag(\hat{B}_0, \cdots, \hat{B}_{T-1}, 0).
\end{aligned}
\end{equation}
An LTV state feedback controller for system~\eqref{eq:LTV_system} is parameterized by $\uu = \KK\xx$ with $\KK\in \mathcal{L}_{TV}^{T, n_u \times n_x}$ and $u_t = \sum_{i=0}^t K^{t,t-i} x_i$ for $t = 0, \cdots, T-1$. Plugging $\uu = \KK \xx$ into~\eqref{eq:open_dyn}, the closed-loop dynamics can be described as 
\begin{equation}
\xx = Z (\hat{\sA} + \hat{\sB} \KK) \xx + \bfeta
\end{equation}
from which we can derive the mapping from the perturbation $\bfeta$ to the closed-loop states $\xx$ and inputs $\uu$ as
\begin{equation} \label{eq:K_mapping}
\begin{bmatrix}
\xx \\ \uu 
\end{bmatrix} = \begin{bmatrix}
(I - Z (\hat{\sA} + \hat{\sB} \KK))^{-1} \\
\KK (I - Z (\hat{\sA} + \hat{\sB} \KK))^{-1}
\end{bmatrix} \bfeta.
\end{equation}
Since $Z$ is a block-down-shifting operator, the matrix inverse in~\eqref{eq:K_mapping} exists and the mapping is well-defined. The maps from $\bfeta$ to $(\xx, \uu)$ in~\eqref{eq:K_mapping} have a block-lower-triangular structure~\eqref{eq:BLT}. We call such maps closed-loop \emph{system responses} and denote them by $\Phix \in \mathcal{L}_{TV}^{T, n_x \times n_x}$, $\Phiu \in \mathcal{L}_{TV}^{T, n_u \times n_x}$ following~\cite{anderson2019system} such that 
\begin{equation} \label{eq:system_response}
	\begin{bmatrix}
	\xx \\ \uu 
	\end{bmatrix} = \begin{bmatrix}
	\Phix \\
	\Phiu
	\end{bmatrix} \bfeta.
\end{equation}
The following theorem characterizes all achievable closed-loop system responses $(\Phix, \Phiu)$ for system~\eqref{eq:LTV_system} under an LTV state feedback controller $\KK$.
\begin{theorem}~\cite[Theorem 2.1]{anderson2019system} 
	\label{thm:SLS}
	Over the horizon $ t = 0, 1, \cdots, T$, for the system dynamics~\eqref{eq:LTV_system} with the block-lower-triangular state feedback control law $\KK \in \mathcal{L}_{TV}^{T, n_u \times n_x}$ defining the control action as $\uu = \KK \xx$, we have:
	\begin{enumerate}
		\item The affine subspace defined by 
		\begin{equation} \label{eq:affine_constr}
		\begin{bmatrix} I - Z \hat{\sA} & -Z \hat{\sB} \end{bmatrix} 
		\begin{bmatrix} \Phix \\ \Phiu \end{bmatrix} = I, \ \Phix, \Phiu \in \mathcal{L}_{TV}
		\end{equation}
		parameterizes all possible system responses~\eqref{eq:system_response}.
		\item For any block-lower-triangular matrices $\lbrace \Phix, \Phiu \rbrace \in \mathcal{L}_{TV}$ satisfying~\eqref{eq:affine_constr}, the controller $\KK = \Phiu \Phix^{-1} \in \mathcal{L}_{TV}$ achieves the desired responses~\eqref{eq:system_response}. 
	\end{enumerate}
\end{theorem}
Theorem~\ref{thm:SLS} establishes the equivalence between $(\Phix, \Phiu)$ and $\KK$ through the affine constraint~\eqref{eq:affine_constr}, and allows us to optimize over the system responses $(\Phix, \Phiu)$ directly in place of $\KK$. In robust SLS MPC~\cite{chen2022robust}, the structure of the robust OCP with the system response parameterization of $\pi$ is exploited for uncertainty over-approximation.

\subsection{Uncertainty over-approximation}
For systems with model uncertainties, the perturbation $\bfeta$ in~\eqref{eq:open_dyn} is dependent on both the uncertainties and the controller $\KK$ to be designed. To show this, note that for a non-delay LTV system with model uncertainty
\begin{equation} \label{eq:uncertain_LTV}
x_{t+1} = \hat{A}_t x_t + \hat{B}_t u_t + \Delta_{A,t} x_t + \Delta_{B,t} u_t + w_t,
\end{equation}
where $w_t \in\mathbb{R}^{n_x}$ is an exogenous disturbance process, the perturbation to the nominal dynamics at time $t$ is given by $\eta_t = \Delta_{A,t} x_t + \Delta_{B,t} u_t + w_t$. Let 
\begin{equation} \label{eq:w_stack}
\ww = [x_0^\top \ w_0^\top \ \cdots \ w_{T-1}^\top]^\top,
\end{equation}
and
\begin{equation}
\begin{aligned}
&\mathbf{\DDelta_A} = \blkdiag(\Delta_{A,0}, \cdots, \Delta_{A,T-1}, 0), \quad \mathbf{\DDelta_B} = \blkdiag(\Delta_{B,0}, \cdots, \Delta_{B,T-1}, 0).
\end{aligned}
\end{equation}
Then for the uncertain system~\eqref{eq:uncertain_LTV}, we have
\begin{equation} 
\bfeta = Z \begin{bmatrix}
\DDelta_A & \DDelta_B 
\end{bmatrix} \begin{bmatrix}
\xx \\ \uu
\end{bmatrix} + \ww.
\end{equation}
By Theorem~\ref{thm:SLS}, under a state feedback controller $\uu = \KK \xx$, the value of perturbation $\bfeta$ is uniquely defined by the following equation
\begin{equation} \label{eq:eta_dynamics}
\bfeta = Z \begin{bmatrix}
\DDelta_A & \DDelta_B 
\end{bmatrix} \begin{bmatrix}
\Phix \\ \Phiu
\end{bmatrix} \bfeta + \ww,
\end{equation}
where the uncertainty parameters $(\DDelta_A, \DDelta_B, \ww)$ and the controller $(\Phix, \Phiu)$ jointly decide the realization of $\bfeta$. In fact, the value of $\bfeta$ is uniquely given by
\begin{equation}\label{eq:value_of_eta}
\bfeta = (I -Z \begin{bmatrix}
\DDelta_A & \DDelta_B 
\end{bmatrix} \begin{bmatrix}
\Phix \\ \Phiu
\end{bmatrix})^{-1} \ww,
\end{equation}
where the matrix inversion exists due to the block-down-shifting operator $Z$. 

Being exact, the characterization of $\bfeta$ in~\eqref{eq:eta_dynamics} is too complex to use for solving the robust OCP, let alone~\eqref{eq:value_of_eta}. In robust SLS MPC, the actual perturbation process $\bfeta$ is over-approximated by a surrogate additive disturbance $\bfSigma \tildeww$ where 
\begin{equation} \label{eq:tildeww}
\tildeww = [x_0^\top \ \tilde{w}_0^\top \ \cdots \ \tilde{w}_{T-1}^\top]^\top
\end{equation}
is a virtual disturbance signal with unit norm-bounded components $\lVert \tilde{w}_t \rVert_\infty \leq 1$, and $\bfSigma \in \mathcal{L}_{TV}^{T, n_x \times n_x}$ is a filter that transforms the virtual disturbance signal $\tildeww$ to over-approximate $\bfeta$ with minimal conservatism. 
\begin{example}[Norm-ball over-approximation]
\label{eg:1}
As an example, in~\cite{chen2021level} the filter $\bfSigma$ is parameterized by $\bfSigma = \blkdiag(I, \sigma_1 I,\cdots, \sigma_{T-1}I)$ where $\sigma_t >0$ such that $\bfSigma \tildeww$ represents a sequence of $\ell_\infty$ norm-bounded disturbances with varying radii $\sigma_t$. Under this parameterization, $\bfSigma \tildeww$ is a valid over-approximation of the perturbation $\bfeta$ if $\lVert \eta_t \rVert_\infty \leq \sigma_t$ for all possible realizations of uncertainty $(\Delta_{A,t}, \Delta_{B,t})$ and $w_t$. 
\end{example}

When the filtered disturbance $\bfSigma \tildeww$ can realize all values of the actual perturbation process $\bfeta$, it suffices to consider the surrogate uncertain dynamics with only additive disturbances $\xx = \hat{\sA} \xx + \hat{\sB} \uu + \bfSigma \tildeww$ to solve the robust OCP. The over-approximation procedure for $\bfeta$ under polytopic uncertainty is shown in~\cite{chen2022robust} where a set of convex constraints on the controller and filter parameters are proposed to guarantee $\bfSigma \tildeww$ is a valid surrogate disturbance. In the next section, we illustrate the difficulty of applying this uncertainty over-approximation scheme to time-delay systems~\eqref{eq:delay_system} and address it by a novel controller design.

%% file: delaySLSMPC.tex
% time delay SLS MPC
\label{sec:delay_effects}
The aforementioned SLS-based scheme of uncertainty over-approximation can be overly conservative when applied on time-delay systems. To see this, we first stack the delayed states $x_{-n_a:-1}$ and inputs $u_{-n_b:-1}$ as  
\begin{equation}
\xx^{-} = [x_{-n_a}^\top \ \cdots \ x_{-1}^\top]^\top, \ \uu^{-} = [u_{-n_b}^\top \ \cdots \ u_{-1}^\top]^\top.
\end{equation}
Note that both $\xx^-$ and $\uu^-$ are known in the robust OCP~\eqref{eq:robustOCP} and are therefore vectors of constants. The dynamics of the uncertain time-delay system in~\eqref{eq:robustOCP} over horizon $T$ can be written as
\begin{equation} \label{eq:delay_representation}
\begin{aligned}
\xx &= Z \hat\sA \xx + Z \hat{\sB} \uu + \underbrace{Z \hat{\sA}^- \xx^- + Z \hat{\sB}^- \uu^-}_{\bd} + \underbrace{Z \DDelta_A \xx + Z \DDelta_B \uu + Z \DDelta_A^- \xx^- + Z \DDelta_B^- \uu^-+ \ww}_{\ddelta} \\
& := Z \hat\sA \xx + Z \hat{\sB} \uu + \bd + \ddelta
\end{aligned}
\end{equation}
where $\xx, \uu, \ww$ defined from~\eqref{eq:x_u} and~\eqref{eq:w_stack} are variables representing future $x_t, u_t, w_t$ in the prediction horizon. The block matrices $(\hat{\sA}, \hat{\sB}, \hat{\sA}^-, \hat{\sB}^-)$ for the time-delay system are given in Appendix~\ref{app:block_matrix} together with their uncertain counterparts $(\DDelta_A, \DDelta_B, \DDelta_A^-, \DDelta_B^-)$. Such block matrices definitions are used for the rest of the paper~\footnote{With a slight abuse of notation, $\hat{\sA}, \hat{\sB}, \DDelta_A, \DDelta_B$ were used in Section~\ref{sec:SLS_MPC} with different definitions. For the rest of the paper, these block matrices refer to those defined in Appendix~\ref{app:block_matrix} such that~\eqref{eq:delay_representation} holds.}.

Note that $\hat{\sA}, \hat{\sB}$ are block-lower-triangular, and $\bd, \ddelta\in \mathbb{R}^{(T+1) n_x}$ defined in~\eqref{eq:delay_representation} capture the effects of delay and model uncertainty on the states $x_{0:T}$. The vector $\bd$ is constant and represents the known effects of delay on future states $\xx$ in an additive manner, while $\ddelta$ lumps the uncertainty-induced perturbation to the nominal dynamics and is dependent on future states $\xx$ and inputs $\uu$. When $n_a = n_b = 0$, $\dd, \DDelta_A^{-}, \DDelta_B^{-}$ vanish and~\eqref{eq:delay_representation} recovers the system dynamics with no delay. In this case, our proposed robust MPC method reduces to~\cite{chen2022robust}.

\subsection{Conservative over-approximation due to delay}
\begin{figure}
\centering
\includegraphics[width = 0.4 \columnwidth]{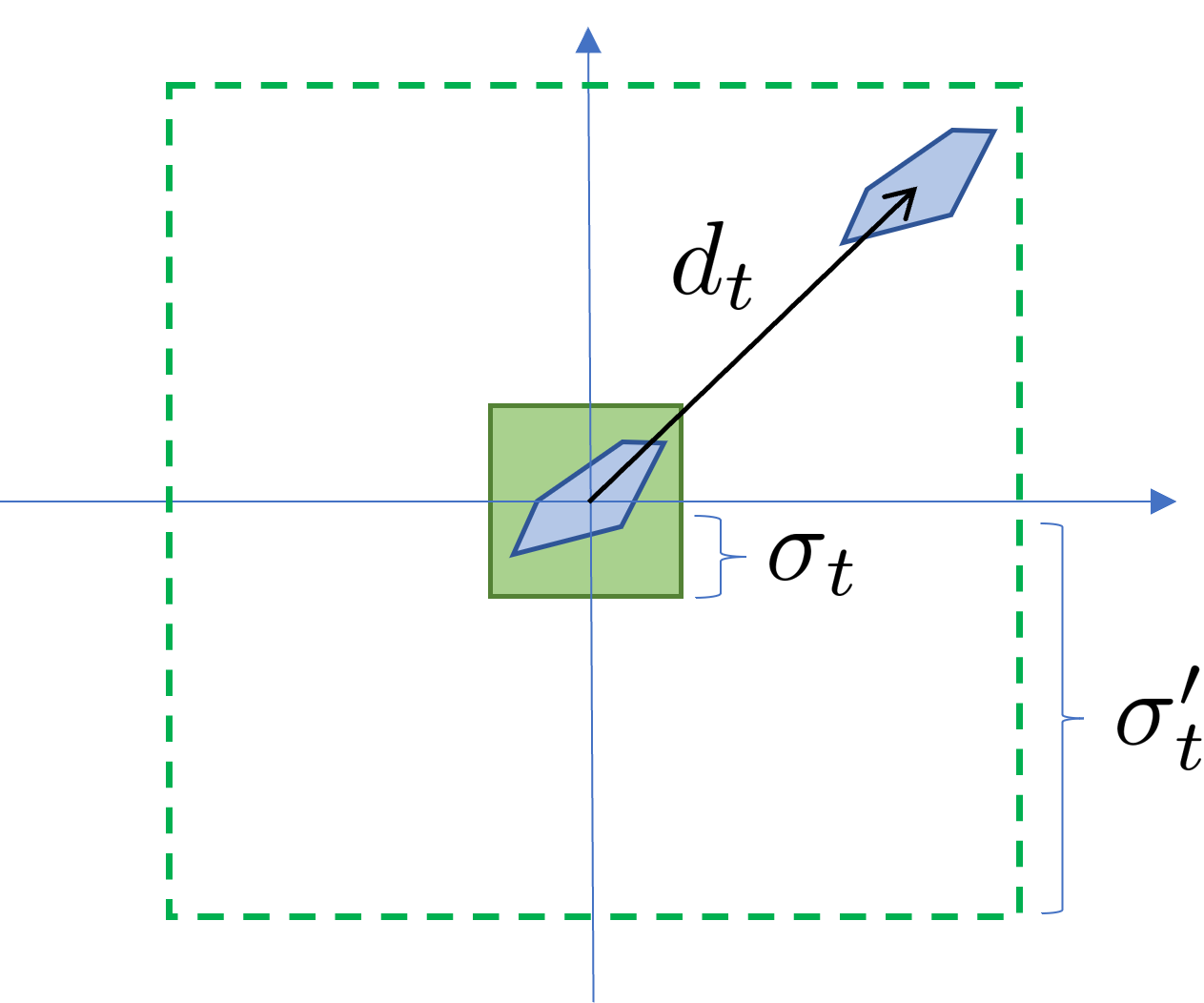}
\caption{The $\ell_\infty$ norm ball over-approximation (green box at the origin) of the uncertainty-induced perturbation $\delta_t$ (blue polytope at the origin) has radius $\sigma_t$, while it becomes $\sigma_t^\prime$ (see box with dashed line) in order to bound the off-set perturbation $d_t + \delta_t$ due to the time-delay effects $d_t$.  }
\label{fig:overapprox}
\end{figure}

To apply robust SLS MPC, we need to over-approximate the perturbation $\bfeta$ (see~\eqref{eq:open_dyn}) to the nominal dynamics by a surrogate disturbance $\bfSigma \tildeww$. For the time-delay system~\eqref{eq:delay_representation}, this indicates treating $\bfeta = \bd + \ddelta$ and bounding $\bfeta$ by the filtered disturbance $\bfSigma \tildeww$. However, this over-approximation can be conservative since the time-delay effects $\bd$ and the uncertainty-induced perturbation $\ddelta$ may differ in scale. For example, with a non-zero delay $(\xx^-, \uu^-)$, the entries in $\dd$ can be large while $\ddelta$ still remains small when the model uncertainties $(\Delta_{A, -n_a:0}, \Delta_{B,-n_b:0}, \sigma_w)$ are close to zero. In this case, the time-delay effects $\bd$ add a non-trivial off-set to the uncertainty-induced perturbation $\ddelta$, and $\bd + \ddelta$ requires significantly larger bounding sets than $\ddelta$ as shown in Figure~\ref{fig:overapprox}. Motivated by this challenge, we propose a feedback controller that acts on both the states $\xx$ and the time-delay effects $\bd$ in order to obtain tighter uncertainty over-approximations. 

\begin{remark}
In robust SLS MPC, the perturbation is chosen as $\bfeta = \bd + \ddelta$ for the time-delay system~\eqref{eq:delay_representation}. In this case, it is reasonable to over-approximate $\bfeta$ by $\bd + \bfSigma \tildeww$ instead of $\bfSigma \tildeww$ since $\bd$ is already known. However, this leads to a non-convex robust OCP formulation in the system responses $(\Phix, \Phiu)$ and the filter $\bfSigma$, in which case the terms $\Phix \bd$ and $\Phiu \bd$ prevent us from convexifying the robust OCP through change of variables (see Section~\ref{sec:change_variable} for details).   
\end{remark}

\subsection{Feedback controller design}
\label{sec:feedback_controller}
We consider only over-approximating the uncertainty-induced perturbation $\ddelta$ by a filtered disturbance $\bfSigma \tildeww$ while handling the known effects of delay $\bd$ separately. Define 
\begin{equation} \label{eq:hh}
	\hh = (I - Z\hat{\sA})^{-1} \bd,  \ \tildexx = \xx - \hh,
\end{equation}
and we denote the components in $\hh$ by $\hh = [h_0^\top \ \cdots \ h_{T}^\top]^\top$. It follows from~\eqref{eq:delay_representation} that
\begin{equation} \label{eq:tilde_system}
\tildexx = Z\hat{\sA} \tildexx + Z \hat{\sB} \uu + \ddelta.
\end{equation}
Then, we can apply Theorem~\ref{thm:SLS} to system~\eqref{eq:tilde_system} with the transformed states $\tildexx$, and obtain that the feedback controller $\uu = \KK \tildexx = \KK(\xx - \hh)$ achieves the following closed-loop system responses
\begin{equation} \label{eq:h_response}
\begin{bmatrix}
\tildexx \\ \uu 
\end{bmatrix} = \begin{bmatrix}
\Phix \\
\Phiu
\end{bmatrix} \ddelta \Rightarrow 	
\begin{bmatrix}
\xx \\ \uu 
\end{bmatrix} = \begin{bmatrix}
\Phix \\
\Phiu
\end{bmatrix} \ddelta + 
\begin{bmatrix}
\hh \\ 0
\end{bmatrix}.
\end{equation}
The controller $\uu = \KK( \xx - \hh)$ applies feedback not only on the future states $x_{0:T}$ but also on the past states and inputs encoded in $\hh$. By integrating the time-delay effects into the transformed states $\tildexx$, the time-delay system~\eqref{eq:tilde_system} now only contains the uncertainty-induced perturbation $\ddelta$ and has the same representation as the non-delay system~\eqref{eq:open_dyn} which is amenable to robust SLS MPC. 

Importantly, variables $\tildexx$ from~\eqref{eq:tilde_system} and $\xx$ from~\eqref{eq:open_dyn} representing future states over the horizon $T$ have the same dimension $(T+1)\times n_x$ which is independent of the delay horizon $n_a$ or $n_b$. This means that the optimization variables introduced in robust SLS MPC for the time-delay system is the same as in the non-delay case. In the next section, we present our solution to the robust OCP~\eqref{eq:robustOCP} using the LTV state-feedback controller $\uu = \KK( \xx - \hh)$.

%Indeed, we can equivalently rewrite~\eqref{eq:hh} as
%\begin{equation} \label{eq:h_definition}
%\hh = Z \hat{\sA} \hh + \bd.
%\end{equation}
%Since the first component of $\bd$ is $0$ which follows from~\eqref{eq:delay_representation}, $\hh$ represents the nominal trajectory $x_{0:T}$ of the time-delay system~\eqref{eq:delay_system} with initial condition $x_0 = 0$ under no actuation, i.e., $u_t = 0$ for $0 \leq t \leq T-1$. 

%% file: SLSMethod.tex
% SLS delay MPC
\label{sec:time_delay_MPC}
In this section, we propose a SLS-based robust MPC method for controlling the uncertain time-delay system~\eqref{eq:delay_system} based on $\uu = \KK (\xx - \hh)$ and the transformed dynamics~\eqref{eq:tilde_system}. As shown in Section~\ref{sec:SLS_MPC}, robust SLS MPC consists of two steps: a) over-approximating the actual perturbation process $\ddelta$ by a surrogate additive disturbance, and b) synthesizing a robust controller based on the surrogate uncertain dynamics.

\subsection{Over-approximation of perturbation $\ddelta$}
Similar to~\eqref{eq:eta_dynamics}, the value of $\ddelta$ for time-delay systems under the controller $\uu = \KK (\xx - \hh)$ is uniquely defined by
\begin{equation} \label{eq:delta_dynamics}
\begin{aligned}
\ddelta &= Z \DDelta_A \Phix \ddelta + Z \DDelta_A \hh + Z \DDelta_B \Phiu \ddelta + Z \DDelta_A^- \xx^- + Z \DDelta_B^- \uu^- + \ww
\end{aligned}
\end{equation}
which follows from~\eqref{eq:delay_representation} and~\eqref{eq:h_response}. Denote the components in $\ddelta$ as $\ddelta = [x_0^\top \ \delta_0^\top \ \cdots \ \delta_{T-1}^\top]^\top$. The value of $\ddelta$ is jointly decided by the uncertainty and the feedback controller applied, and is therefore also uncertain. 
%Due to the block-downshift operator $Z$, it follows from~\eqref{eq:delta_dynamics} that the value of $\delta_t$ only depends on $x_0$ and $\delta_{0:t-1}$. 
Our goal is to over-approximate $\ddelta$ by a filtered signal $\bfSigma \tildeww$ where $\tildeww$ is given in~\eqref{eq:tildeww} with $\lVert \tilde{w}_t \rVert_\infty \leq 1$. In other words, for all possible values of $\ddelta$, we want to guarantee that there exists $\tildeww$ such that $\ddelta = \bfSigma \tildeww$ and $\lVert \tilde{w}_t \rVert_\infty \leq 1$ for $ 0 \leq t \leq T-1$. We denote the unit norm-bounded constraint on $\tilde{w}_t$ as $\tilde{w}_t \in \mathcal{W}_{\tildeww} = \{ w \in \mathbb{R}^{n_x} \mid \lVert w \rVert_\infty \leq 1\}$. Since the value of $\ddelta$ is uniquely defined by~\eqref{eq:delta_dynamics}, it is equivalent to showing that
\begin{equation}\label{eq:tildeww_dyn}
\begin{aligned}
\bfSigma \tildeww &= Z \DDelta_A \Phix \bfSigma \tildeww + Z \DDelta_A \hh + Z \DDelta_B \Phiu \bfSigma \tildeww + Z \DDelta_A^- \xx^- + Z \DDelta_B^- \uu^- + \ww
\end{aligned}
\end{equation}
is robustly feasible with $\tilde{w}_t \in \mathcal{W}_{\tildeww}$ for all possible realization of $(\Delta_{A,0:n_a}, \Delta_{B,0:n_b}) \in \mathcal{P}$ and $w_t \in \mathcal{W}$. We provide sufficient conditions for~\eqref{eq:tildeww_dyn} to hold robustly through the following steps. 

\subsubsection{Change of variable}
\label{sec:change_variable}
To avoid non-convexity in our formulation, we first do the change of variable
\begin{equation}\label{eq:change_of_variable}
\tildePhix = \Phix \bfSigma, \quad \tildePhiu = \Phiu \bfSigma.
\end{equation}
Under the condition that $\bfSigma$ is invertible, by~\cite[Theorem 1]{chen2021level} all achievable $(\tildePhix, \tildePhiu)$ are directly parameterized by
\begin{equation} \label{eq:tilde_affine_constr}
\begin{bmatrix} I - Z \hat{\sA} & -Z \hat{\sB} \end{bmatrix} 
\begin{bmatrix} \tildePhix \\ \tildePhiu \end{bmatrix} = \bfSigma, \ \tildePhix, \tildePhiu \in \mathcal{L}_{TV},
\end{equation}
where $(\tildePhix, \tildePhiu)$ can be interpreted as the system responses mapping $\tildeww$ to $(\tildexx, \uu)$ under the controller $\uu = \KK \tildexx$ for the system $\tildexx = Z \hat{\sA} \tildexx + Z \hat{\sB} \uu + \bfSigma \tildeww$. Then, in an optimization problem, searching over $(\Phix, \Phiu, \bfSigma)$ with the affine constraint~\eqref{eq:affine_constr} is equivalent to searching $(\tildePhix, \tildePhiu, \bfSigma)$ with constraint~\eqref{eq:tilde_affine_constr} (see \cite[Remark 1]{chen2021level} for details).

\subsubsection{Parameterization of the filter $\bfSigma$}
\label{sec:sigma_params}
The filter $\bfSigma \in\mathcal{L}_{TV}^{T, n_x \times n_x}$ has the block-lower-triangular structure~\eqref{eq:BLT}, but its diagonal blocks are specially parameterized. We set $\Sigma^{0,0} = I$ such that the first component of the filtered disturbance $\bfSigma \tildeww$ is $x_0$. The other block matrices on the diagonal of $\bfSigma$ are parameterized as $\Sigma^{t,0} = \diag(q_{t-1})$ where $q_{t-1} \in \mathbb{R}^{n_x}$ and $q_{t-1} > 0$ for $t = 1, \cdots, T$. By this parameterization, $\Sigma^{t,0}$ are themselves diagonal matrices with positive entries and therefore $\bfSigma$ is invertible. More importantly, such parameterization allows us to formulate convex sufficient conditions in $(\tildePhix, \tildePhiu, \bfSigma)$ to over-approximate $\ddelta$ by $\bfSigma \tildeww$ as shown in Section~\ref{sec:over_approx_delta}.

\begin{example}[Hyperrectangle over-approximation]
	When the lower-triangular blocks $\Sigma^{i, j}$ for $j > 0$ are enforced zero, the filtered disturbance $\bfSigma \tildeww$ with the above parameterization of $\bfSigma$ represents a sequence of disturbances bounded by hyperrectangles. This provides us more flexibility in bounding the actual perturbation $\ddelta$ than the norm-ball over-approximation shown in Example~\ref{eg:1}.
\end{example}

\subsubsection{Constraint simplification}
\label{sec:constr_simplify}
We now simplify our notations to represent the equality constraints in~\eqref{eq:tildeww_dyn}. Decompose the filter as $\bfSigma = \bfSigma_{diag} + \bfSigma_{sub}$ where $\bfSigma_{diag} \in \mathcal{L}_{TV}^T$ is the block-diagonal matrix that contains the matrices $\Sigma^{t,0}$, $t = 0, \cdots, T$ which are on the diagonal of $\bfSigma$, and $\bfSigma_{sub}\in \mathcal{L}_{TV}^T$ contains the rest lower-triangular blocks. Under the change of variable~\eqref{eq:change_of_variable}, we can rewrite the equality constraints in \eqref{eq:tildeww_dyn} as
\begin{equation}\label{eq:tilde_equality}
\begin{aligned}
(\bfSigma_{diag} + \bfSigma_{sub}) \tildeww &= Z \DDelta_A \tildePhiu \tildeww + Z \DDelta_A \hh + Z \DDelta_B \tildePhix \tildeww + Z \DDelta_A^- \xx^- + Z \DDelta_B^- \uu^- + \ww. 
\end{aligned}
\end{equation}
Then we group the terms in~\eqref{eq:tilde_equality} as
\begin{equation} \label{eq:derive_1}
\begin{aligned}
\bfSigma_{diag} \tildeww & = \underbrace{(Z \DDelta_A \tildePhix +  Z \DDelta_B \tildePhiu - \bfSigma_{sub})}_{\CC} \tildeww + Z\DDelta_A \hh + Z \DDelta_A^- \xx^- + Z \DDelta_B^- \uu^- + \ww
\end{aligned}
\end{equation}
where we define $\CC = Z \DDelta_A \tildePhix +  Z \DDelta_B \tildePhiu - \bfSigma_{sub}$ and $\CC \in \mathcal{L}_{TV}^{T}$. Let $\CC_0 = \CC(:,0)$ denote the first block column of $\CC$ and $\CC_T = \CC(:, 1:T)$ denote the rest block columns. We have $\CC \tildeww = \CC_0 x_0 + \CC_T \tilde{w}_{0:T-1}$ where $\tilde{w}_{0:T-1} = [\tilde{w}_0^\top \ \cdots  \ \tilde{w}_{T-1}^\top]^\top$ is the stack of all virtual disturbances. It follows from~\eqref{eq:derive_1} that
\begin{equation} \label{eq:separation}
\begin{aligned}
\bfSigma_{diag} \tildeww & = \CC_T \tilde{w}_{0:T-1} + \ww + \underbrace{\CC_0 x_0 + Z\DDelta_A \hh + Z \DDelta_A^- \xx^- + Z \DDelta_B^- \uu^-}_{\vv}  \\
& := \CC_T \tilde{w}_{0:T-1} + \vv + \ww
\end{aligned}
\end{equation}
where $\vv$ as defined above encodes the effects of the initial condition and the time delay. We observe that once the model uncertainties $(\Delta_{A,0:n_a}, \Delta_{B,0:n_b})$ are fixed, the entries in $\CC_T$ and $\vv$ are linear in the design parameters $(\tildePhix, \tildePhiu, \bfSigma)$, and vice versa. Since $\CC_T$ is a truncation of the block-lower-triangular matrix $\CC$, we refer to the matrix blocks in $\CC_T$ using their indices in $\CC$, i.e., by $C^{t,t-i}$, and similarly for $\bfSigma_{diag}$. By construction, the diagonal blocks of $\CC$ and the first component of $\vv$ are zero. Therefore, we denote the entries in $\vv$ as $\vv = [0 \ v_1^\top \ \cdots \ v_{T-1}^\top]^\top$ and we have $C^{t,0} = 0$ for $0 \leq t \leq T$. 

\subsubsection{Over-approximation constraints}
\label{sec:over_approx_delta}
By writing down the equality constraints in~\eqref{eq:separation} row-wise and plugging in the parameterization of $\bfSigma$, we have
\begin{equation} \label{eq:simple_constr}
\begin{aligned}
I x_0 &= x_0, \\
\diag(q_0) \tilde{w}_0 &= v_0 + w_0, \\
\diag(q_t)\tilde{w}_t & = \sum_{i=1}^{t} C^{t+1, i} \tilde{w}_{i-1} + v_t + w_t, \\
\lVert \tilde{w}_t \rVert_\infty &\leq 1, \quad t = 1, \cdots, T-1.
\end{aligned}
\end{equation}
Note that $v_t$ and $C^{i,j}$ are in fact functions of the model uncertainty and the controller. 
%We first consider how to guarantee~\eqref{eq:simple_constr} has a feasible solution $\tildeww$ satisfying $\lVert \tilde{w}_t \rVert_\infty \leq 1$ for all possible values of the disturbance $\ww$ with $\lVert w_t \rVert_\infty \leq \sigma_w$. 
One important feature of~\eqref{eq:simple_constr} is that the value of $\tilde{w}_t$ only depends on $\tilde{w}_{0:t-1}$, and this allows us to synthesize a robust feasible solution to~\eqref{eq:simple_constr} sequentially as follows. 

In~\eqref{eq:simple_constr}, the first constraint $Ix_0 = x_0$  holds by construction. The second constraint corresponding to $t = 0$ holds only if 
\begin{equation}
\lVert \diag(q_0)^{-1} (v_0 + w_0 ) \rVert_\infty \leq 1, \ \forall  \lVert w_0 \rVert_\infty \leq \sigma_w,
\end{equation}
which is equivalent to 
\begin{equation} \label{eq:temp_1}
\begin{aligned}
&\lvert e_i^\top (v_0 + w_0) \rvert \leq q_{0, i}, 1 \leq i \leq n_x, \forall \lVert w_0 \rVert_\infty \leq \sigma_w \Leftrightarrow \lvert e_i^\top v_0 \rvert + \sigma_w \leq q_{0,i}, 1 \leq i \leq n_x,
\end{aligned}
\end{equation}
where $q_{0,i}$ denotes the $i$-th entry of $q_0$, and $e_i$ is the $i$-th standard basis. Constraint~\eqref{eq:temp_1} is obtained by the triangle inequality and H\"older's inequality, and guarantees the existence of $\tilde{w}_0^* \in \mathcal{W}_{\tildeww}$ such that $\diag(\tilde{w}_0^*) = v_0 + w_0$ for all possible values of $w_0$. To further robustify the constraint against the underlying model uncertainty, we note that constraint~\eqref{eq:temp_1} is convex in $v_0$, and $\vv$ is an affine function of the model uncertainty parameters $(\Delta_{A,0:n_a}, \Delta_{B,0:n_b})$ when the system responses $(\tildePhix, \tildePhiu)$ are fixed. Therefore, the left-hand side (LHS) of~\eqref{eq:temp_1} is convex in $(\Delta_{A,0:n_a}, \Delta_{B,0:n_b})$. Using the fact that the maximum of a convex function over a polytope domain is achieved at the polytope vertices~\cite{boyd2004convex}, we can tighten constraint~\eqref{eq:temp_1} as
\begin{equation} \label{eq:temp_4}
\begin{aligned}
& \lvert e_i^\top v_0 \rvert + \sigma_w \leq q_{0,i}, \forall (\Delta_{A,0:n_a}, \Delta_{B,0:n_b}) \in \mathcal{P} \\
\Leftrightarrow &\lvert e_i^\top v_0 \rvert + \sigma_w \leq q_{0,i}, \forall (\Delta_{A,0:n_a}, \Delta_{B,0:n_b}) \in \vertex(\mathcal{P})
\end{aligned}
\end{equation}
for $i = 1, \cdots, n_x$, where $\vertex(\cdot)$ denotes the set of vertices of the polytopic uncertainty set $\mathcal{P}$. Now constraint~\eqref{eq:temp_4} guarantees $\diag(q_0) \tilde{w}_0 = v_0 + w_0$ is robustly feasible with $\tilde{w}_0 \in \mathcal{W}_{\tildeww}$ for all possible model uncertainty and additive disturbances. Furthermore, since $\vv$ is affine in $(\tildePhix,\tildePhiu, \bfSigma)$ when $(\Delta_{A,0:n_a}, \Delta_{B,0:n_b})$ are fixed, \eqref{eq:temp_4} is convex in the design parameters $(\tildePhix,\tildePhiu, \bfSigma)$.

Now we consider the constraint for $t =1$:
\begin{equation} \label{eq:temp_2}
\diag(q_1) \tilde{w}_{1} = C^{2,1} \tilde{w}_0^* + v_1 + w_1,
\end{equation}
where we have applied the solution $\tilde{w}_0^*$ synthesized from the previous step. Although the exact value of $\tilde{w}_0^*$ depends on $w_0$ and is unknown, with the information that $\lVert \tilde{w}_0^* \rVert_\infty \leq 1$, we can treat $\tilde{w}_0^*$ as a norm-bounded disturbance. Applying the same technique, we have that
\begin{equation} \label{eq:temp_3}
\begin{aligned}
& \lvert e_i^\top v_1 \rvert +  \lVert e_i^\top C^{2,1} \rVert_1 + \sigma_w \leq q_{1,i}, \ i = 1, \cdots, n_x,\ \forall (\Delta_{A,0:n_a}, \Delta_{B,0:n_b}) \in \vertex(\mathcal{P})
\end{aligned}
\end{equation} 
guarantees constraint~\eqref{eq:temp_2} is robustly feasible with $\tilde{w}_1 \in \mathcal{W}_{\tildeww}$. Repeat this process up to $t = T-1$, we have that the following constraints on $\CC_T$ and $\vv$ 
\begin{equation}\label{eq:over_approx_constr}
\begin{aligned}
&\lvert e_i^\top v_1 \rvert +  \lVert e_i^\top C^{2,1} \rVert_1 + \sigma_w \leq q_{1,i},  \\
& \lvert e_i^\top v_t \rvert + \sum_{i=1}^t \lVert e_i^\top C^{t+1, i} \rVert_1 + \sigma_w \leq q_{t, i}, \\
& \forall (\Delta_{A,0:n_a}, \Delta_{B,0:n_b}) \in \vertex(\mathcal{P}), \ i = 1, \cdots, n_x,  \ t= 1, \cdots, T-1,
\end{aligned}
\end{equation}
guarantee that~\eqref{eq:simple_constr} is robustly feasible for the considered polytopic model uncertainty and additive disturbances, and $\bfSigma \tildeww$ is a valid over-approximation of the uncertainty-induced perturbation $\ddelta$. Again, since the model uncertainty parameters are fixed, constraints~\eqref{eq:over_approx_constr} are convex in $(\tildePhix, \tildePhiu, \bfSigma)$. 

\subsection{Robust OCP formulation}
Under the uncertainty over-approximation constraint~\eqref{eq:over_approx_constr}, we can apply the surrogate dynamics 
\begin{equation} \label{eq:surrogate_dyn}
\begin{aligned}
\tildexx =Z \hat{\sA} \tildexx + Z \hat{\sB} \uu + \bfSigma \tildeww
\end{aligned}
\end{equation}
to solve the robust OCP~\eqref{eq:robustOCP} where $\tildexx = \xx - \hh$. Recall that the affine constraint~\eqref{eq:tilde_affine_constr} parameterizes all achievable system responses 
\begin{equation} \label{eq:temp_6}
	\tildexx = \tildePhix \tildeww, \quad \uu = \tildePhiu \tildeww
\end{equation}
for system~\eqref{eq:surrogate_dyn} under the feedback controller $\uu = \KK\tildexx = \KK(\xx - \hh)$. Next, we apply these relationships to solve the robust OCP~\eqref{eq:robustOCP} with robust constraint satisfaction guarantees. 

Assume that the polyhedral state constraint consists of $n_\mathcal{X}$ linear constraints, i.e., $\mathcal{X} = \{x \mid F_\mathcal{X}(i,:) x \leq b_\mathcal{X}(i), i = 1, \cdots, n_{\mathcal{X}}\}$, and denote the set of the linear constraint parameters as $\facet(\mathcal{X}) = \{(f,b) \vert f = F_\mathcal{X}(i,:), b =  b_\mathcal{X}(i), i = 1,\cdots, n_{\mathcal{X}} \}$. Then, based on the surrogate uncertain dynamics~\eqref{eq:surrogate_dyn} and the achievable system responses~\eqref{eq:temp_6}, a robust state constraint in the robust OCP~\eqref{eq:robustOCP} can be written as
\begin{equation} \label{eq:temp_5}
\begin{aligned}
f^\top x_t = f^\top(\tilde{x}_t + h_t) = f^\top (\tilde{\Phi}_x^{t,t} x_0 + \sum_{i=1}^{t}\tilde{\Phi}_x^{t,t-i} \tilde{w}_{i-1} + h_t ) \leq b, \ \forall \tilde{w}_j \in \mathcal{W}_{\tildeww}, \ j = 0, \cdots, t-1.
\end{aligned}
\end{equation}
By applying the H\"older's inequality to constraint~\eqref{eq:temp_5} for $0 \leq t \leq T$, we can tighten all the state constraints in the robust OCP~\eqref{eq:robustOCP} as
\begin{equation} \label{eq:state_tightening}
\begin{aligned}
&f^\top (\tilde{\Phi}_x^{t,t} x_0 + h_t) + \sum_{i=1}^{t} \lVert f^\top \tilde{\Phi}_x^{t,t-i} \rVert_1 \leq b, \ \forall (f,b)\in \facet(\mathcal{X}), \ t = 0, \cdots, T-1.
\end{aligned}
\end{equation}
Similarly, we tighten the terminal constraint $x_T \in \mathcal{X}_T$ as 
\begin{equation} \label{eq:terminal_tightening}
\begin{aligned}
& f^\top (\tilde{\Phi}_x^{T,T} x_0 + h_T) + \sum_{i=1}^{T} \lVert f^\top \tilde{\Phi}_x^{T,T-i} \rVert_1 \leq b, \ \forall (f,b)\in \facet(\mathcal{X}_T),
\end{aligned}
\end{equation}
and tighten the control input constraints $u_t \in \mathcal{U}$ as
\begin{equation} \label{eq:control_tightening}
\begin{aligned}
&f^\top (\tilde{\Phi}_u^{t,t} x_0 + h_t) + \sum_{i=1}^{t} \lVert f^\top \tilde{\Phi}_u^{t,t-i} \rVert_1 \leq b, \ \forall (f,b)\in \facet(\mathcal{U}), \ t =0 , \cdots, T-1.
\end{aligned}
\end{equation}
The tightened constraints~\eqref{eq:state_tightening}, \eqref{eq:terminal_tightening}, \eqref{eq:control_tightening} are convex in $(\tildePhix, \tildePhiu)$. When applied in conjunction with the uncertainty over-approximation constraint~\eqref{eq:over_approx_constr}, constraints \eqref{eq:state_tightening}, \eqref{eq:terminal_tightening}, \eqref{eq:control_tightening} guarantee that the synthesized controller $\uu = \KK (\xx - \hh)$ with $\KK= \tildePhiu \tildePhix^{-1}$ is feasible for the robust OCP~\eqref{eq:robustOCP}. We formally summarize the proposed method in the following theorem. 

\begin{theorem} \label{thm:guarantee}
Consider the following convex OCP:
\begin{equation} \label{eq:convex_inner_approx}
\begin{aligned}
\underset{\tildePhix, \tildePhiu, \bfSigma}{\textrm{minimize}} & \quad J_T(\tildePhix(:,0) x_0 + \hh, \tildePhiu(:,0)x_0)  \\
\textrm{subject to} & \quad \textrm{affine constraint~\eqref{eq:tilde_affine_constr}} \\
& \quad \textrm{uncertainty over-approximation constraint~\eqref{eq:over_approx_constr}} \\
& \quad \textrm{tightened constraints~\eqref{eq:state_tightening}, \eqref{eq:terminal_tightening}, \eqref{eq:control_tightening}} \\
&\quad x_{-n_a:0}, u_{-n_b:-1} \textrm{ are known}
\end{aligned}
\end{equation}
where $\bfSigma \in \mathcal{L}_{TV}^{T, n_x \times n_x}$ is parameterized in Section~\ref{sec:sigma_params}, and the parameters $\CC, \vv$ used in constraint~\eqref{eq:over_approx_constr} are defined in Section~\ref{sec:constr_simplify}. Then, for any feasible solution $(\tildePhix, \tildePhiu, \bfSigma)$ of problem~\eqref{eq:convex_inner_approx}, the feedback controller $\uu = \KK(\xx - \hh)$, where $\hh$ is defined in~\eqref{eq:hh} and $\KK= \tildePhiu \tildePhix^{-1}$, is feasible for the robust OCP~\eqref{eq:robustOCP}. 
\end{theorem}

The proof of Theorem~\ref{thm:guarantee} directly follows from our derivation of the constraints in problem~\eqref{eq:convex_inner_approx} in this section. In robust SLS MPC, we apply a nominal quadratic cost function $J_T(\cdot)$ in problem~\eqref{eq:convex_inner_approx} where $\tildePhix(:,0) x_0 + \hh$ and $\tildePhiu(:,0)x_0$ represent the nominal states and control inputs for the surrogate dynamical system~\eqref{eq:surrogate_dyn}, respectively, by setting $\tilde{w}_t = 0$. Since all constraints in~\eqref{eq:convex_inner_approx} can be formulated as linear constraints, the robust OCP~\eqref{eq:convex_inner_approx} is a convex QP. We note that the dimensions of the decision variables $(\tildePhix, \tildePhiu, \bfSigma)$ in~\eqref{eq:convex_inner_approx} are decided by the system dimensions $(n_x, n_u)$ and the prediction horizon $T$ while being independent of the delay horizon $(n_a, n_b)$.

%% file: new_simulation.tex
% new simulation

% Simulation
\label{sec:simulation}
%We demonstrate our proposed robust MPC method on a truck-trailer control problem from~\cite{shi2009delay}. The truck-trailer model has a $3$-dimensional state $x = [x_1 \ x_2 \ x_3]^\top$ with its entries denoting the angle difference between the truck and the trailer, the angle of the trailer, and the position of the rear end of the trailer sequentially. The $1$-dimensional control input $u$ denotes the steering angle. Using the same model parameters and the discretization procedure with sampling time $0.1$ sec in~\cite{shi2009delay}, we have an uncertain discrete-time delayed system:
We test the effectiveness and scalability of the proposed method through numerical examples. All the simulation is implemented in MATLAB R2019b with YALMIP~\cite{lofberg2004yalmip} and MOSEK~\cite{mosek} on an Intel i7-6700K CPU.

\subsection{3D system}
We demonstrate our proposed robust MPC method on a 3-dimensional time-delay system from~\cite{shi2009delay}:
\begin{equation}\label{eq:truck_trailer}
\begin{aligned} 
&x(k+1) = \begin{bmatrix}
1.0509 & 0 & 0 \\
-0.0509 & 1 & 0 \\
0.0509 \alpha(k) & -0.4 \alpha(k) & 1
\end{bmatrix} x(k) + \\
&\begin{bmatrix}
0.0218 & 0 & 0 \\
-0.0218 & 0 & 0 \\
0.0218 \alpha(k) & 0 & 0
\end{bmatrix}
x(k-3)+ 
\begin{bmatrix}
-0.1429 \\ 0 \\ 0
\end{bmatrix} u(k) + w(k)
\end{aligned}
\end{equation}
where $\alpha(k) \in [1, 1.5915]$ is a time-varying uncertain parameter. We consider the same control input constraints $\lvert u(k) \rvert \leq \pi$ as in~\cite{shi2009delay}, but additionally we consider state constraints $\mathcal{X} = \{x \in \mathbb{R}^3 \mid \lvert x_1 \rvert \leq \frac{2}{3}\pi, \lvert x_2 \rvert \leq 2\pi, \lvert x_3 \rvert \leq 15\}$ and additive disturbances $w(k)$ bounded by $\lVert w(k) \rVert_\infty \leq 0.05$. 
%System~\eqref{eq:truck_trailer} can be easily transformed to the form~\eqref{eq:delay_system} where the model uncertainty set $\mathcal{P}$ has two vertices corresponding to $\alpha(k) = 1$ and $\alpha(k) = 1.5915$. 

\begin{figure}
	\centering
	\begin{subfigure}{0.45 \columnwidth}
		\includegraphics[width = \textwidth]{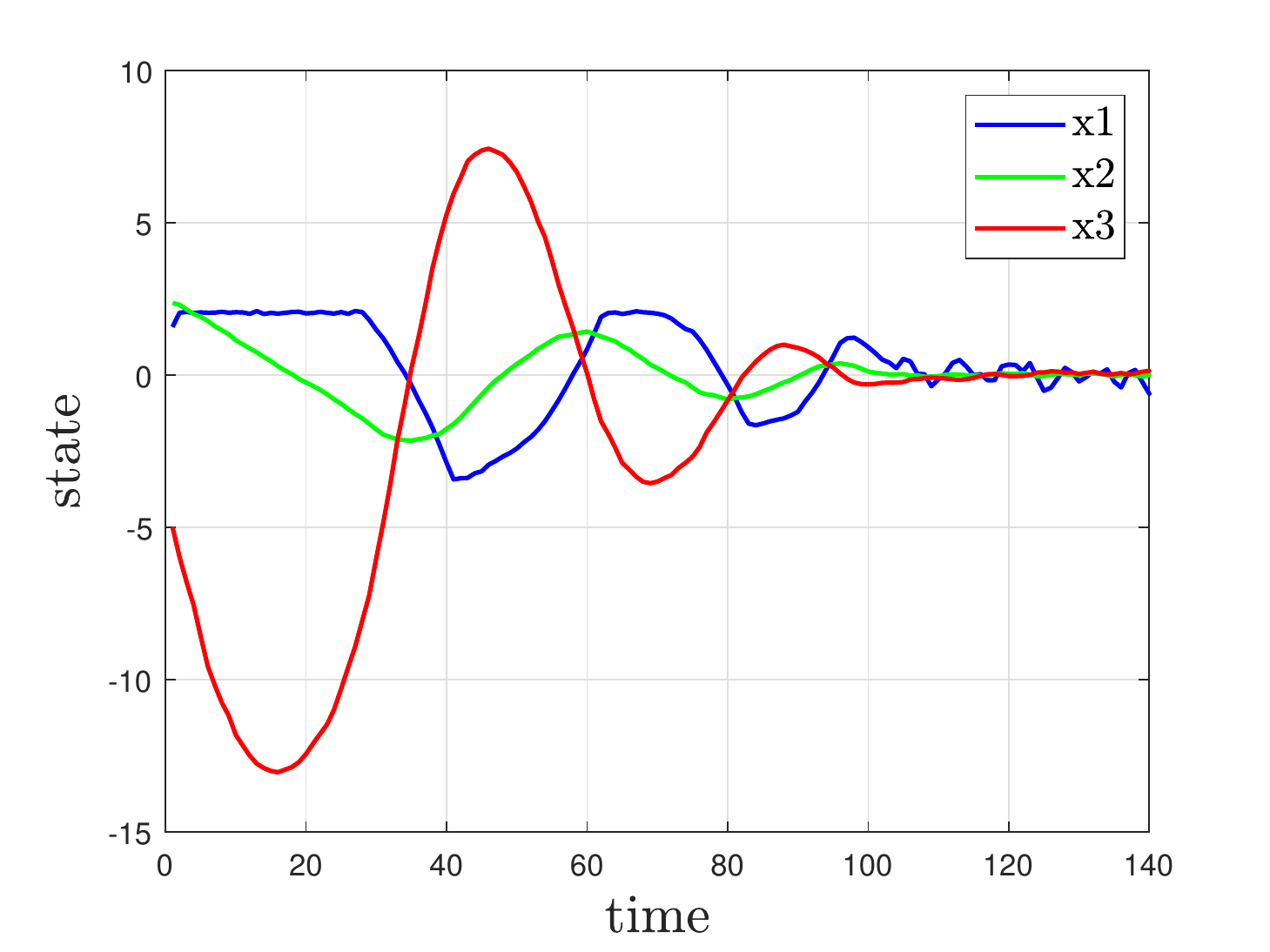}
	\end{subfigure}
	\hfill
	\begin{subfigure}{0.45 \columnwidth}
		\includegraphics[width = \textwidth]{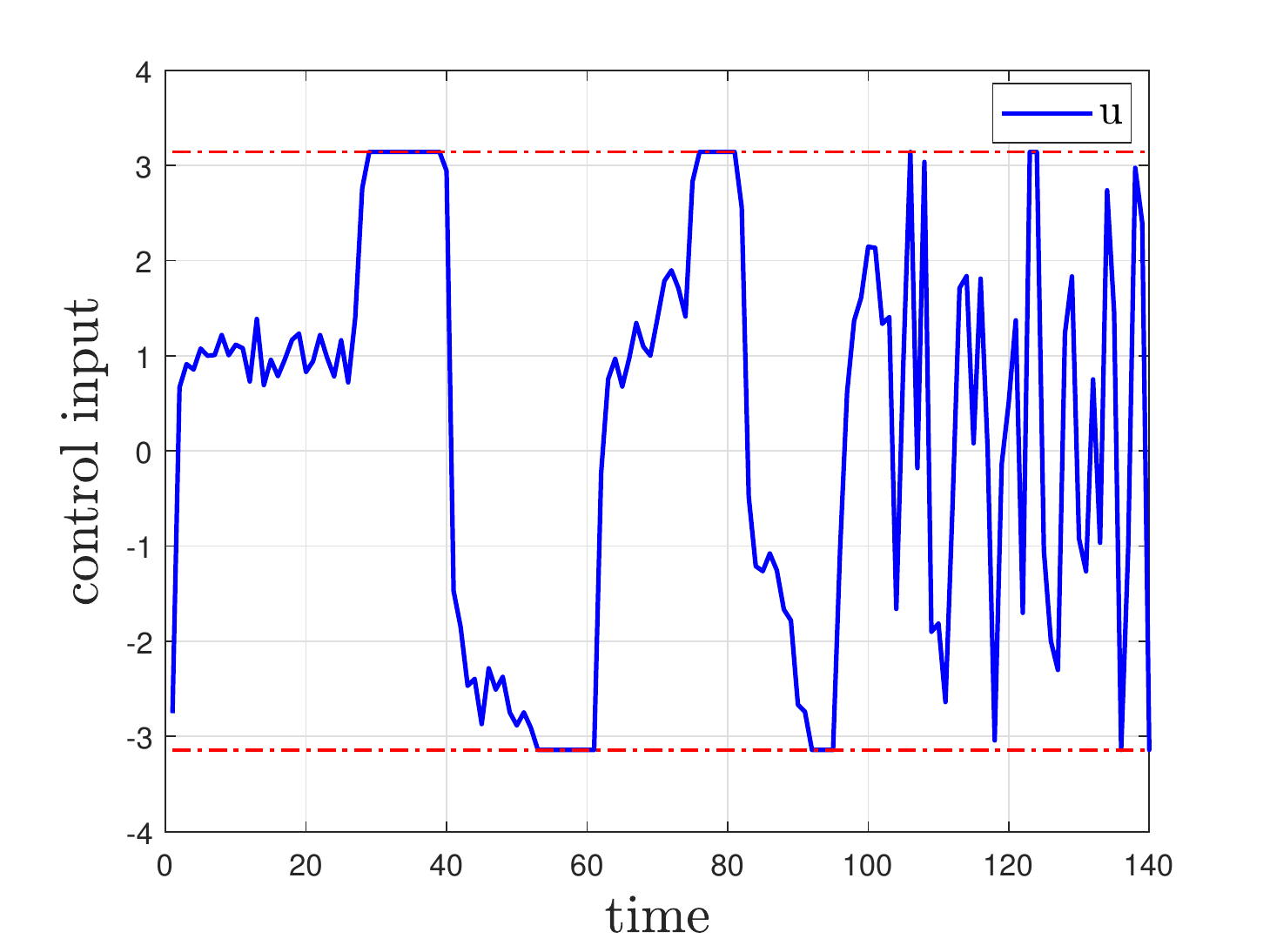}
	\end{subfigure}
	\caption{Closed-loop trajectory of the states (left) and control inputs (right) of the uncertain system~\eqref{eq:truck_trailer} under our proposed robust MPC controller. Additive disturbances are injected.  }
	\label{fig:noisy}
\end{figure}

With the initial condition $x(0) = [0.5 \pi \ 0.75 \pi \ -5]^\top, x(-1) = x(-2) = x(-3) = 0$, we apply our method with horizon $T = 6$ to evolve system~\eqref{eq:truck_trailer} in closed-loop. No terminal constraint is applied in~\eqref{eq:convex_inner_approx}, and the cost function $J_T$ is chosen as a quadratic function in the nominal states and control inputs
\begin{equation}\label{eq:nominal_cost}
\begin{aligned}
J_T(\hat{x}_{0:T}, \hat{u}_{0:T-1}) = \sum_{t=0}^{T-n_a} \hat{x}_t^\top Q \hat{x}_t + \sum_{t=T-n_a + 1}^{T} \hat{x}_T^\top Q_T \hat{x}_T + \sum_{t=0}^{T-1} \hat{u}_t R \hat{u}_t
\end{aligned}
\end{equation}
where $Q = I$, $R = 0.01$ and $Q_T = 100I$. In the simulation, the uncertainty parameter $\alpha(k)$ is uniformly sampled from the interval $[1, 1.5915]$ and $w(k)$ is uniformly sampled from the box $\lVert w(k) \rVert_\infty \leq 0.05$ at each time instant. Figure~\ref{fig:noisy} shows that our method guarantees the satisfaction of constraints for the closed-loop system in the presence of both model uncertainty and additive disturbances.

\begin{table}
	\centering
	\begin{tabular}{cc}
		\toprule
		$(n_a, n_b, T)$ & Solver time/s \\ \midrule
		$(8,4,13)$ & $0.2052 $ \\ \midrule
		$(16,8,21)$ & $0.3404$ \\ \midrule
		$(24,12,29)$ & $0.9372$\\ \midrule
		$(32,16,37)$ & $1.7020$\\ \midrule
		$(40,20,45)$ & $2.8175$ \\ 
		\bottomrule
	\end{tabular}
	\quad
	\begin{tabular}{cc}
		\toprule
		$(n_a, n_b, T)$ & Solver time/s \\ \midrule
		$(0,0,13)$ & $0.1332 $ \\ \midrule
		$(0,0,21)$ & $0.2078$ \\ \midrule
		$(0,0,29)$ & $0.3476$\\ \midrule
		$(0,0,37)$ & $0.4870$\\ \midrule
		$(0,0,45)$ & $0.8468$ \\ 
		\bottomrule
	\end{tabular}
	\caption{Solver time of the robust OCP~\eqref{eq:convex_inner_approx} with different ranges of delay horizon $(n_a, n_b)$ and prediction horizon $T$. For each $(n_a, n_b, T)$, the average solver time over $20$ randomly generated time-delay systems is reported.}
	\label{tab:solver_time}
\end{table}

\subsection{Scalability test}
We demonstrate the scalability of the proposed method with respect to the delay horizon on randomly generated systems in Table~\ref{tab:solver_time}. We fix the state dimension as $n_x = 2$ and input dimension as $n_u = 1$. For a given delay horizon $(n_a, n_b)$ and prediction horizon $T > \max(n_a, n_b)$, we randomly generate dynamics matrices $A_i, 0 \leq i \leq n_a$ and $B_j, 0 \leq j \leq n_b$ whose entries are independently sampled from the normal distribution $\mathcal{N}(0, 0.09)$. State constraints $\mathcal{X} = \mathcal{X}_T = \{ x \in \mathbb{R}^2 \mid \lVert x \rVert_\infty \leq 30\}$ and input constraints $\mathcal{U} = \{u\in \mathbb{R} \mid \lvert u \rvert \leq 5\}$ are enforced, and the nominal quadratic cost function~\eqref{eq:nominal_cost} with $Q = Q_T = I, R = 1$ are considered in the robust OCP~\eqref{eq:convex_inner_approx}. Model uncertainty is introduced as an unknown parameter $\alpha \in [0, 1]$ such that 
\begin{equation*}
	\Delta_{A,i} = \alpha \begin{bmatrix}
	-0.1 & 0 \\ 0 & 0 
	\end{bmatrix} + (1 - \alpha) \begin{bmatrix}
	0.1 & 0 \\ 0 & 0
	\end{bmatrix}, i = 0, \cdots, n_a.
\end{equation*}
The uncertainty on $B_j$ and the additive disturbances are not considered in this example. Finally, the initial condition is fixed as $x_0 = [2.5 \ -2.5]^\top$ and $x_{-i} = 0, u_{-j} = 0$ for all delayed states and inputs.

In Table~\ref{tab:solver_time}, we report the average solver time of the robust OCP~\eqref{eq:convex_inner_approx} for a range of delay and prediction horizons $(n_a, n_b, T)$. 
%We choose $n_b = n_a/2, T = n_a + 5$ and range the state delay horizon $n_a$ from $8$ to $40$. 
The QP~\eqref{eq:convex_inner_approx} is solved by MOSEK~\cite{mosek} on an Intel i7-6700K CPU. Since we do not use any states or inputs augmentation, there is no substantial increase in the solver time of our method as the delay horizon grows large. Indeed, for $(n_a, n_b, T) = (40, 20, 45)$, the common approach that augments the time-delay system as a non-delay LTI system as shown in~\cite{laraba2017linear} gives rise to a system of $(n_a+1) \times n_x + n_b\times n_u = 102$ dimension which is challenging to handle for robust MPC methods with a horizon $T = 45$. Compared with the non-delay problem instances (right half in Table~\ref{tab:solver_time}), our proposed robust SLS MPC approach only suffers from the increase of number of constraints in the QP~\eqref{eq:convex_inner_approx} due to non-zero $(n_a, n_b)$ while sharing the same number of optimization variables.

%% file: conclusion.tex
% conclusion
\label{sec:conclusion}
We proposed an SLS-based robust MPC method for uncertain discrete-time linear systems with time delay. Our method handles the effects of time delay by incorporating them into the feedback controller design, and leverages SLS to bound the perturbation induced by the polytopic model uncertainty and norm-bounded additive disturbances in the dynamics. Our method solves a convex quadratic program online whose number of variables is independent of the delay horizon. 

%% file: new_appendix.tex
% new_appendix
% appendix
\section{Time-delay system representation}
\label{app:block_matrix}
The block matrices in the compact representation of the time-delay system in~\eqref{eq:delay_representation} are explicitly defined below. We take $\hat{\sA}^-$ and $\hat{\sA}$ as an example which gives
\begin{equation*} \label{eq:delay_BLT}
\mathbf{\hat{A}}^{-} = \begin{bmatrix}
\hat{A}_{n_a} & \cdots & \cdots & \hat{A}_1 \\
0 & \hat{A}_{n_a} & \cdots & \hat{A}_2 \\
\vdots & \ddots & \ddots & \vdots \\
\vdots & \ &  \ddots & \hat{A}_{n_a} \\
\mathbf{0} & \cdots & \cdots & \mathbf{0}
\end{bmatrix} \in\mathbb{R}^{(T+1)n_x \times n_a n_x}
\end{equation*} 
where $\mathbf{0} \in \mathbb{R}^{(T+1-n_a)n_x \times n_x}$ is a zero matrix, and 
\begin{equation*} \label{eq:nominal_BLT}
\mathbf{\hat{A}} = \begin{bmatrix}
\hat{A}_0 & \ & \ & \ & \ & \ \\
\vdots & \ddots & \ & \ & \ & \ \\
\hat{A}_{n_a}  & \vdots & \ddots & \ & \ & \ \\
0 & \ddots & \vdots & \ddots & \ & \ \\
\vdots & \ddots & \hat{A}_{n_a} & \cdots & \hat{A}_0 & \ \\
0 & \cdots & 0 & \cdots & \cdots & 0
\end{bmatrix} 
\end{equation*} 
with the dimension $\mathbf{\hat{A}} \in\mathbb{R}^{(T+1)n_x \times (T+1)n_x}$. The block matrices $\hat{\sB}, \hat{\sB}^-$ and the uncertain block matrices $\DDelta_A, \DDelta_B, \DDelta_A^-, \DDelta_B^-$ have the same structure with different different matrices inserted.

%% file: main.bbl
\begin{thebibliography}{10}

\bibitem{kothare1996robust}
M.~V. Kothare, V.~Balakrishnan, and M.~Morari, ``Robust constrained model
  predictive control using linear matrix inequalities,'' {\em Automatica},
  vol.~32, no.~10, pp.~1361--1379, 1996.

\bibitem{langson2004robust}
W.~Langson, I.~Chryssochoos, S.~Rakovi{\'c}, and D.~Q. Mayne, ``Robust model
  predictive control using tubes,'' {\em Automatica}, vol.~40, no.~1,
  pp.~125--133, 2004.

\bibitem{goulart2006optimization}
P.~J. Goulart, E.~C. Kerrigan, and J.~M. Maciejowski, ``Optimization over state
  feedback policies for robust control with constraints,'' {\em Automatica},
  vol.~42, no.~4, pp.~523--533, 2006.

\bibitem{sieber2021system}
J.~Sieber, S.~Bennani, and M.~N. Zeilinger, ``A system level approach to
  tube-based model predictive control,'' {\em IEEE Control Systems Letters},
  vol.~6, pp.~776--781, 2021.

\bibitem{chen2022robust}
S.~Chen, V.~M. Preciado, M.~Morari, and N.~Matni, ``Robust model predictive
  control with polytopic model uncertainty through system level synthesis,''
  {\em arXiv preprint arXiv:2203.11375}, 2022.

\bibitem{fridman2014introduction}
E.~Fridman, {\em Introduction to time-delay systems: Analysis and control}.
\newblock Springer, 2014.

\bibitem{shi2009delay}
Y.-J. Shi, T.-Y. Chai, H.~Wang, and C.-Y. Su, ``Delay-dependent robust model
  predictive control for time-delay systems with input constraints,'' in {\em
  2009 American Control Conference}, pp.~4880--4885, IEEE, 2009.

\bibitem{bououden2016constrained}
S.~Bououden, M.~Chadli, L.~Zhang, and T.~Yang, ``Constrained model predictive
  control for time-varying delay systems: Application to an active car
  suspension,'' {\em International Journal of Control, Automation and Systems},
  vol.~14, no.~1, pp.~51--58, 2016.

\bibitem{ding2007constrained}
B.~Ding and B.~Huang, ``Constrained robust model predictive control for
  time-delay systems with polytopic description,'' {\em International Journal
  of Control}, vol.~80, no.~4, pp.~509--522, 2007.

\bibitem{laraba2017linear}
M.-T. Laraba, S.~Olaru, and S.-I. Niculescu, ``Linear model predictive control
  and time-delay implications,'' {\em IFAC-PapersOnLine}, vol.~50, no.~1,
  pp.~14406--14411, 2017.

\bibitem{olaru2008predictive}
S.~Olaru and S.-I. Niculescu, ``Predictive control for linear systems with
  delayed input subject to constraints,'' {\em IFAC Proceedings Volumes},
  vol.~41, no.~2, pp.~11208--11213, 2008.

\bibitem{anderson2019system}
J.~Anderson, J.~C. Doyle, S.~H. Low, and N.~Matni, ``System level synthesis,''
  {\em Annual Reviews in Control}, vol.~47, pp.~364--393, 2019.

\bibitem{chen2021level}
S.~Chen, N.~Matni, M.~Morari, and V.~M. Preciado, ``System level
  synthesis-based robust model predictive control through convex inner
  approximation,'' {\em arXiv preprint arXiv:2111.05509}, 2021.

\bibitem{boyd2004convex}
S.~Boyd, S.~P. Boyd, and L.~Vandenberghe, {\em Convex optimization}.
\newblock Cambridge university press, 2004.

\bibitem{lofberg2004yalmip}
J.~Lofberg, ``Yalmip: A toolbox for modeling and optimization in matlab,'' in
  {\em 2004 IEEE international conference on robotics and automation (IEEE Cat.
  No. 04CH37508)}, pp.~284--289, IEEE, 2004.

\bibitem{mosek}
M.~ApS, {\em The MOSEK optimization toolbox for MATLAB manual. Version 9.0.},
  2019.

\end{thebibliography}
